\RequirePackage[hyphens]{url}
\documentclass{codee}
\usepackage{graphicx}
\usepackage[square,numbers]{natbib}
\usepackage{hyperref}
\usepackage{enumitem}
\newtheorem{theorem}{Theorem}[section]

\theoremstyle{definition}

\newtheorem{probl}[theorem]{Problem}
\newtheorem{question}[theorem]{Discussion Question}

\theoremstyle{remark}

\numberwithin{equation}{section}
\usepackage{xcolor}
\usepackage{soul}
\title{ODEs and Mandatory Voting}
\author{Christoph B\"orgers, Natasa Dragovic, Anna Haensch, Arkadz Kirshtein}
\affiliation{Tufts University, Medford, MA}
\author{Lilla Orr}
\affiliation{University of Richmond, Richmond, VA}
\shortauthor{One, Two and Three}
\date{June, 2023}
\dedicatory{}
\submissiondate{June, 2023}

\begin{document}
\maketitle

\begin{abstract}
This paper presents mathematics relevant to the question whether
voting should be mandatory. 
Assuming a static distribution of voters' political beliefs, we model how politicians might adjust their positions to raise their
share of the vote. Various scenarios can be explored using our
app  at \url{https://centrism.streamlit.app/}. 
Abstentions are found to have great impact on the dynamics
of candidates, and in particular to introduce the possibility
of discontinuous jumps in optimal candidate positions. This is
an unusual application of ODEs. We hope that it might help engage
some students who may find it harder to connect with the more 
customary applications from the natural sciences.
\end{abstract}

\section{Introduction}

Mandatory voting has been adopted by over 20 of the world’s democracies,
including Brazil and Australia, and has been discussed in the United States as well \citep{Brennan2014,Craig_22,Brookings_20}.
Mathematical modeling may help understand its effects. To accompany  this article, we have created a  web app hosted at \url{https://centrism.streamlit.app/} that allows students to experiment with problem parameters and outcomes. The app, though built with Python, has a friendly graphical user interface that requires no knowledge of coding. As an intermediate step between the  app and the pure Python scripts, we have also prepared a public Python workbook in Google Colab that walks through the relevant models discussed in this paper, available at \url{https://annahaensch.com/centrism.html}.  This workbook is appropriate for a student who has some  familiarity with coding.  For the curious and Python proficient student the code underlying the web app (as well as sample Python scripts) is also available in the  public GitHub repository \url{https://github.com/annahaensch/Centrism}.  
Note that the web app and the paper have the same general structure but the paper has answers to the discussion questions and additional homework problems for the students. 

The organization of the paper is as follows: in Section~\ref{sec:distribution_voter_beliefs} we introduce the model
of voters' beliefs. We assume that the distribution of voters' beliefs does not change over time. In Section~\ref{sec:mandatory_voting} we start with a scenario where every voter casts a vote, which is what
{\em nearly} happens when voting is mandatory. Under this assumption, we explore which political candidate wins when (1) candidates' positions are fixed, or (2)  candidates are given time to optimize their positions.  
Section~\ref{sec:optional_voting} has the same outline as Section~\ref{sec:mandatory_voting}, except now allowing for the possibility that voters choose not to vote when neither political candidate is sufficiently appealing. In Section~\ref{sec:discountinuities} we take a closer look at what happens from an ODE perspective when voters are allowed to abstain, specifically looking at possible discontinuities in the system. Finally, in Section~\ref{sec:conclusion} we  summarize our results.

Throughout the 
paper we  provide {\em in-class discussion questions} (with
our own thoughts on those questions), and {\em homework problems}
to complement the class discussion.

\section{Distribution of Voter Beliefs}\label{sec:distribution_voter_beliefs}

\begin{figure}
\centering
  \includegraphics[width=4.5in]{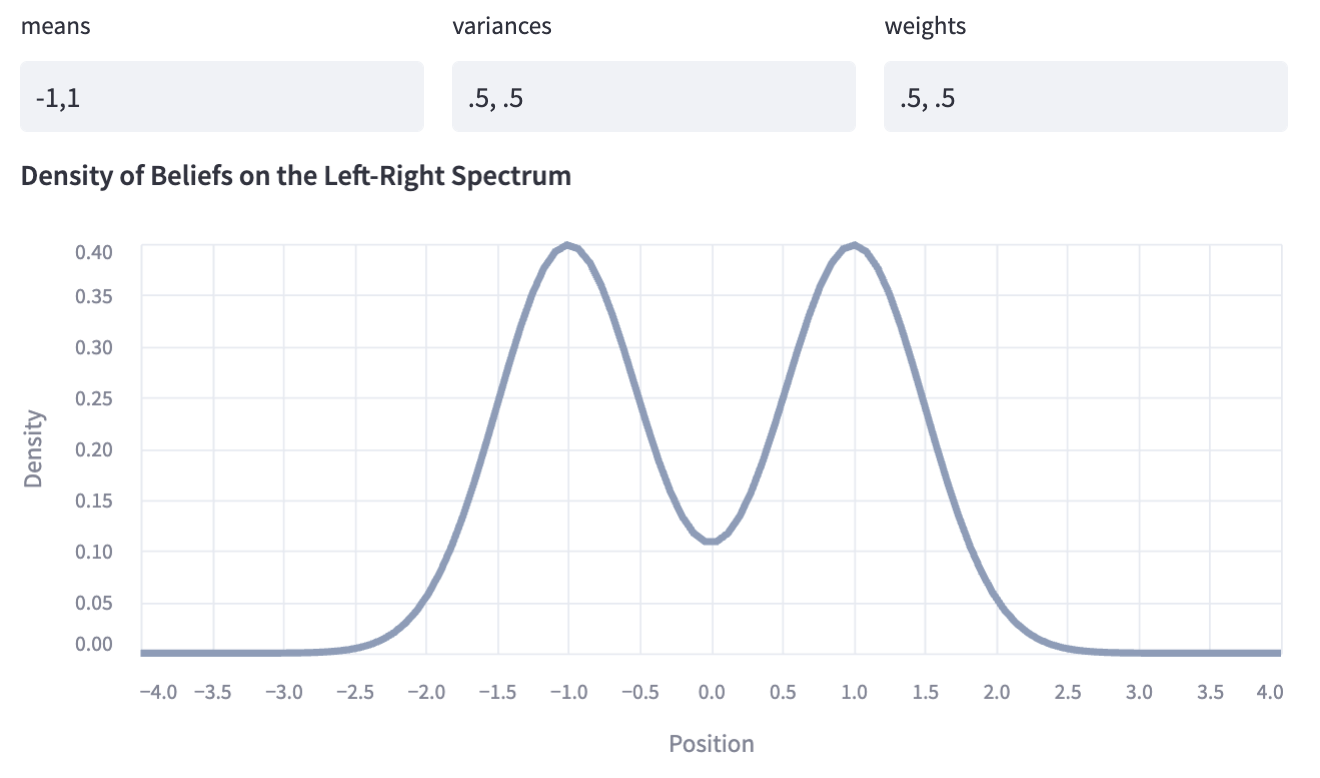}
  \caption{Example of a distribution of voter beliefs}
  \label{fig:ex_voter_beliefs}
\end{figure}

Throughout the paper, we assume that the political candidates and voters' political beliefs can be represented on a "left-right" axis. 
We'll start by seeding a population with political beliefs on the left-right spectrum. To keep things simple, we assume that the 
belief distribution is a {\em Gaussian mixture}, that is, it is
described by a weighted average of Gaussian densities with
different means and variances. As an example,
Figure \ref{fig:ex_voter_beliefs} shows the average of two Gaussian probability density
functions, with means $-1$ and $1$ and both variances equal to $0.5$.

In the distribution of Figure \ref{fig:ex_voter_beliefs}, there are two ``camps" of voters,
a ``left-wing" and a ``right-wing" camp. 
Very few people hold the most extreme liberal (less than -3) or extreme conservative views (greater than 3), many people hold views that are fairly liberal (-1.5 to -0.5) or conservative (1.5 to 0.5), and a few people hold moderate views (-0.5 to 0.5). 

\begin{question}
\label{reasonable} 
    Does a weighted average of Gaussian densities seem like a reasonable representation of the views of U.S. voters?
\end{question}

In general, a Gaussian mixture is a density of the form
\begin{align}
    f(x)=\sum_{n=1}^M\omega_m\cdot \frac{1}{\sigma_m\sqrt{2\pi}}e^{-\frac{1}{2}\Big(\frac{x-\mu_m}{\sigma_m}\Big)^2}
\end{align}
with weights $\omega_m \geq 0$ that sum to 1.
In our app, you can enter the means and variances that you'd like to use for each of the Gaussian modes. The default is $M=2$, but the app allows any $M$ between $1$ and $5$.

\section{When Voting is Mandatory}\label{sec:mandatory_voting}

From now on, we will assume that there are two political candidates $L$ and $R$. We denote their positions by $l$ and $r$, and assume 
$l<r$.
In this Section, we assume that voting is mandatory, i.e. everyone votes. As a reminder, countries like Australia and Brazil make
voting mandatory. We will first see who wins the election based only on the initial position of the candidates in Subsection~\ref{subsec:static_mandatory} and then allow the candidates to move in Subsection~\ref{subsec:dyn_mandatory}.

\subsection{Static Candidates}\label{subsec:static_mandatory}

We assume that every person will vote for the candidate whose position is closest to their beliefs. Therefore the  left candidate gets  all of the votes to the left of the midpoint between $l$ and $r$. The vote shares of the left and right candidates, $S_L$ and $S_R$, respectively, are
\begin{equation}
\label{eq:SLSR} 
    S_L=\int_{-\infty}^{\frac{l+r}{2}}f(x)dx \mbox{ ~and~ } S_R=\int_{\frac{l+r}{2}}^\infty f(x)dx 
\end{equation}

\begin{figure}[ht]
\centering
  \includegraphics[width=5.4in]{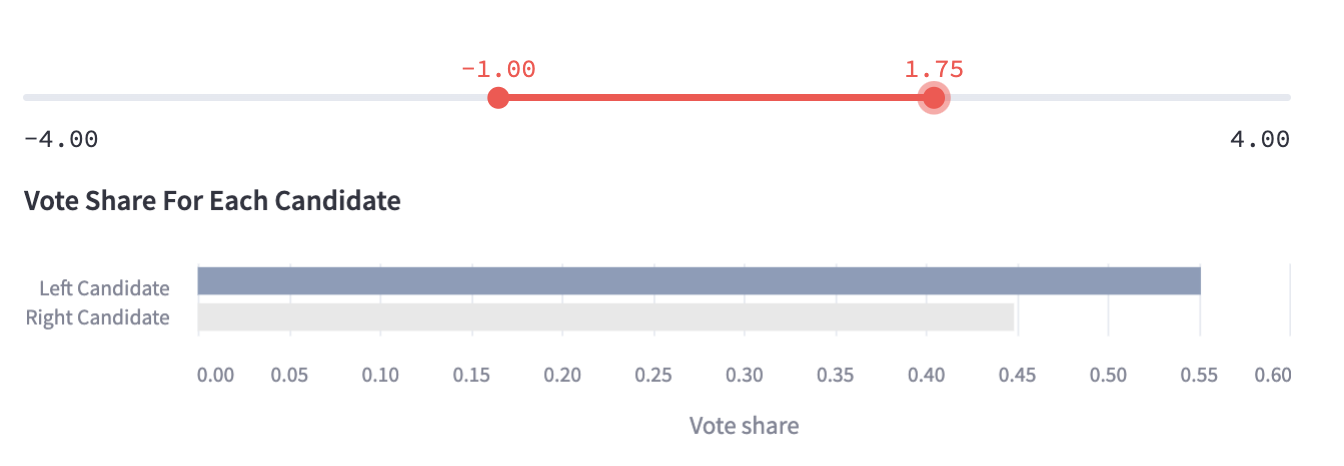}
  \caption{Example of an initial candidate positions}
  \label{fig:ex_initial_candidates}
\end{figure}

\begin{figure}
\centering
  \includegraphics[width=5.4in]{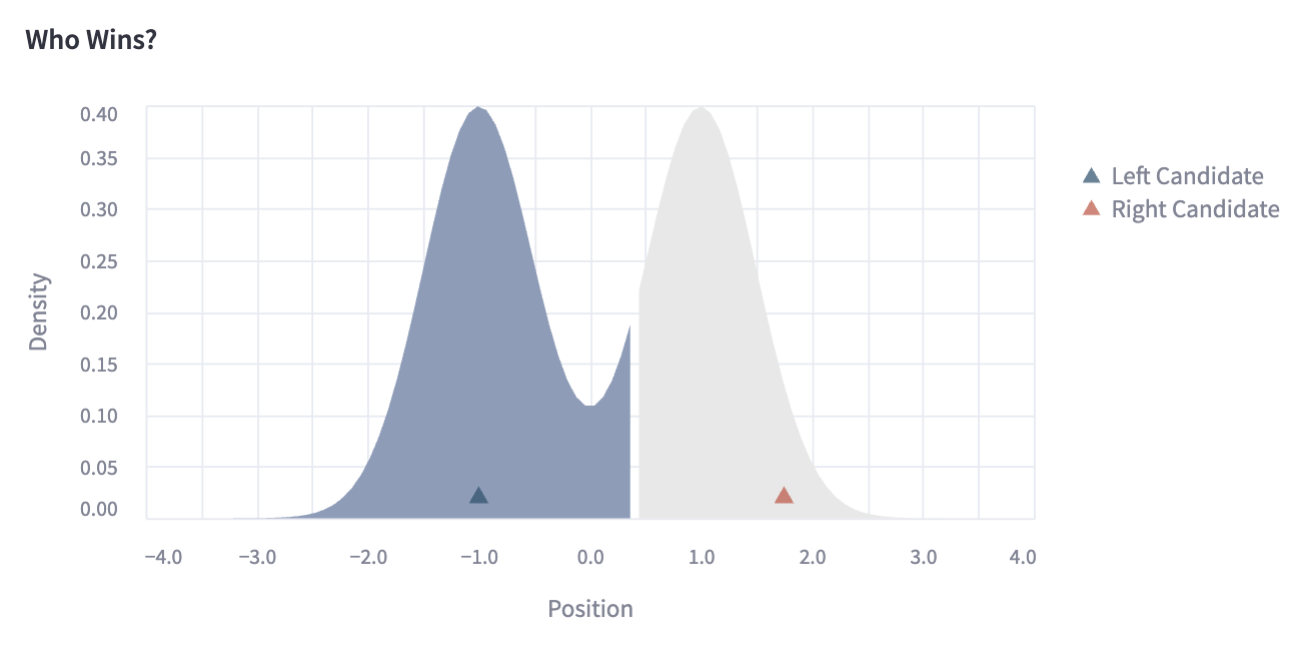}
  \caption{Example of a voter share. Blue shaded area is larger than the gray area, therefore the left candidate wins}
  \label{fig:who_wins_static}
\end{figure}

In our app, you can use the slider to pick the positions of the left and the right candidate; see Figure~\ref{fig:ex_initial_candidates}. 
If $S_L>S_R$, then $L$ wins, and if $S_R>S_L$, $R$ wins; 
see Figure~\ref{fig:who_wins_static}.

\begin{question}
\label{assump} 
    This model is making an assumption about what happens when a country adopts a mandatory voting policy. Please explain this assumption in your own words and discuss possible violations.
\end{question}

\subsection{Dynamic Candidates}\label{subsec:dyn_mandatory}

\begin{figure}
\centering
  \includegraphics[width=5.5in]{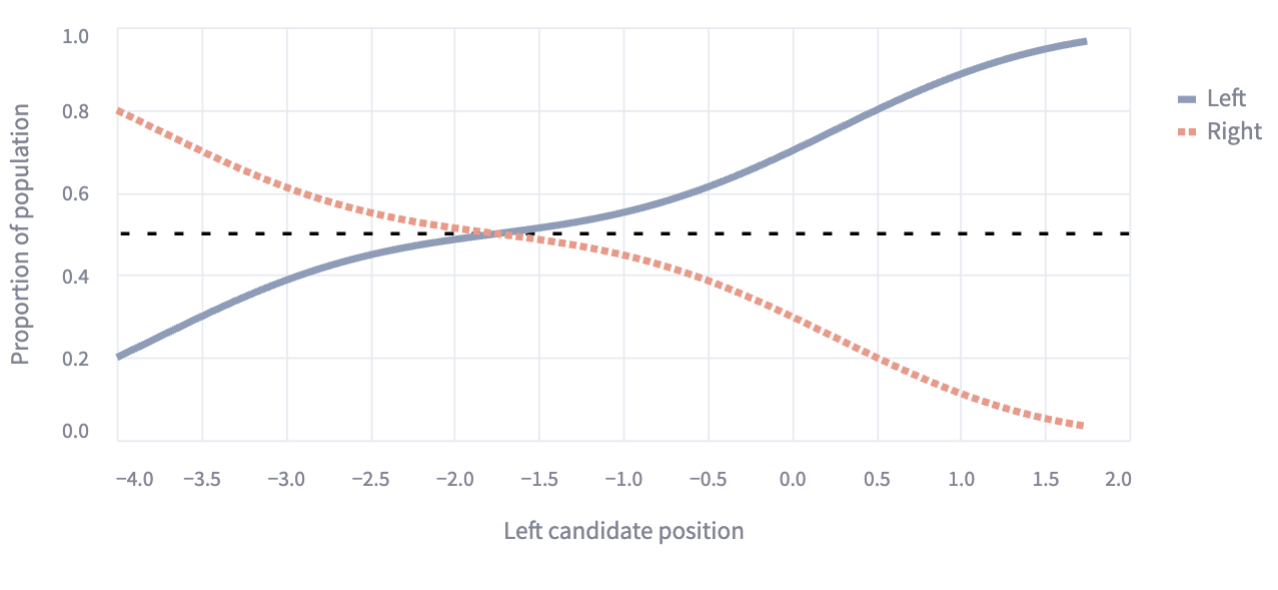}
  \caption{Proportion of population voting for each candidate as a function of left candidate position}
  \label{fig:prop_pop_voting_for_each_candidate}
\end{figure}

If the right candidate has a fixed position, it is  always in the best interest of the left candidate to move closer to the right candidate. To see this more clearly, we compute the vote shares
$S_L$ and $S_R$ as  functions of the left candidate position, $l$, for a fixed $r$;
see Figure \ref{fig:prop_pop_voting_for_each_candidate}. 
As soon as $S_L$ rises above the dashed line, $L$ has more than $50\%$ of the votes and therefore wins. 

\begin{question}
\label{why_so_close} 
    Why does the proportion of the population voting for the left candidate get so close to one, but never quite reach it?
\end{question}

\begin{question}
\label{summ} 
    Do the proportions always sum to one? Why or why not?
\end{question}

\begin{question}
\label{where_to_go} 
    Where does the left candidate have to be on the spectrum in order to win the election?
\end{question}

\begin{figure}
\centering
  \includegraphics[width=5in]{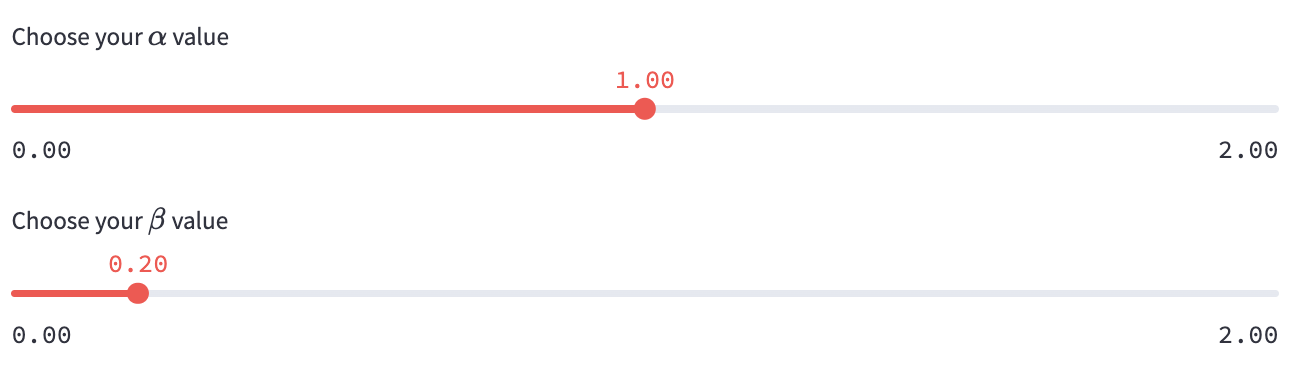}
  \caption{Example of values for $\alpha$ and $\beta$ in our app.}
  \label{fig:example_alpha_beta}
\end{figure}

\begin{figure}
\centering
  \includegraphics[width=5in]{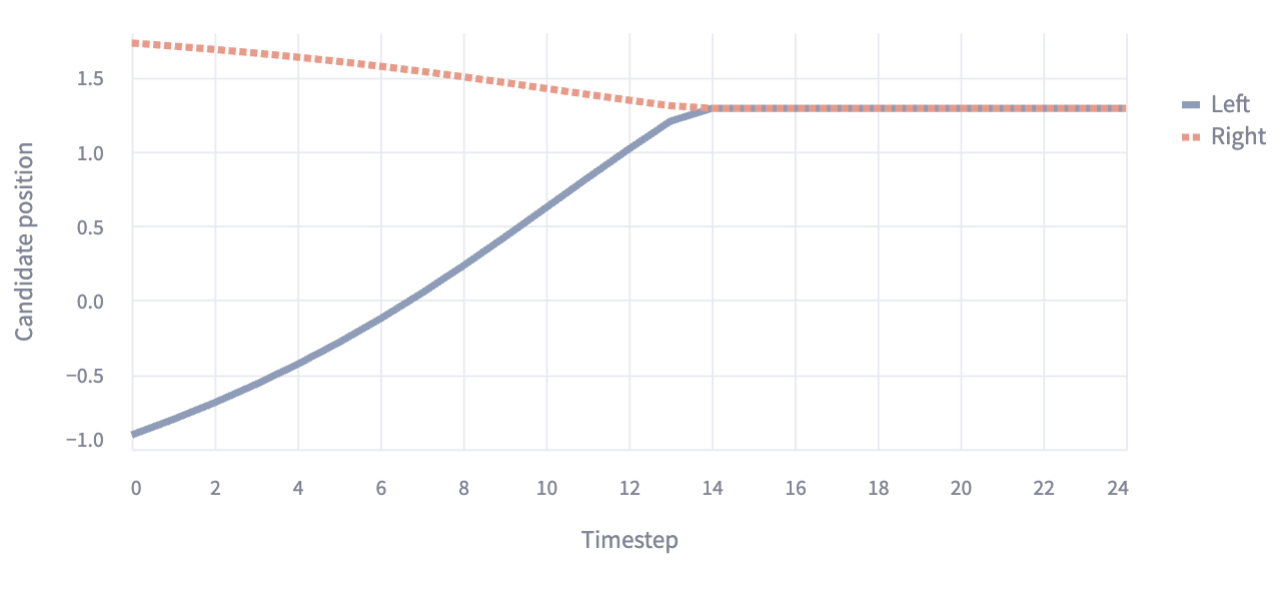}
  \caption{Candidates moving according to steepest ascent }
  \label{fig:cand_moving_steep_ascent}
\end{figure}

\vskip 10pt
From the previous discussion we've seen that candidates might want to change their position on the political spectrum in order to get more votes. Some candidates will do this more eagerly than others. Let's include a measure of \textbf{candidate opportunism} into our model, and the larger this measure is, the more eagerly (i.e. more rapidly) a candidate will move to enlarge their share of the vote.
If both candidates are opportunistic, they might for instance follow a system of ordinary differential equations of the form
\begin{align}
    \frac{ dl}{dt}=\alpha \frac{\partial S_L}{\partial l} \mbox{ ~~and~~ } \frac{ dr}{dt}=\beta \frac{\partial S_R}{\partial r}.
    \label{eq:steepest_ascent}
\end{align}
This is called {\em steepest ascent} --- each candidate
moves in the direction in which their vote share increases, 
given the other candidate's position.
The parameters $\alpha \geq 0$ and $\beta \geq 0$ measure
the degrees of opportunism of $L$ and $R$, respectively.
In our app you can use sliders to pick  values for $\alpha$ and $\beta$. We will pick values 1 and 0.2 for $\alpha$ and $\beta$ respectively; see Figure \ref{fig:example_alpha_beta}.
We then plot the candidates' movements governed by eqs.\ \ref{eq:steepest_ascent}; see Figure \ref{fig:cand_moving_steep_ascent}.

\begin{figure}
\centering
  \includegraphics[width=5in]{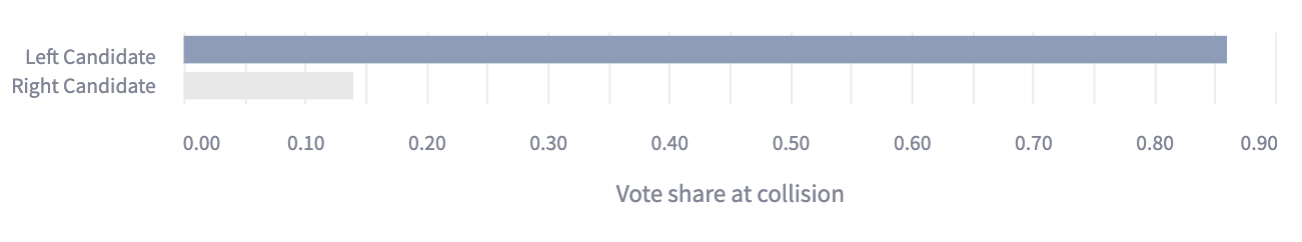}
  \caption{Proportion of population voting for each candidate }
  \label{fig:prop_voting_each_candidate}
\end{figure}

The candidates will eventually meet as some collision point. In the Figure \ref{fig:prop_voting_each_candidate} we show the vote shares for each candidate at this point.

\begin{question}
\label{non_symmetry}
    What happens to the position at which the
    two candidates meet when the left candidate is very opportunistic and the right candidate is not?
\end{question}

\begin{question}
\label{less_eager_wins}
    Is there ever a way for the less eager candidate to win?
\end{question}

\begin{question}
\label{what_if_they_cross} 
    The model proposed here doesn't allow the candidates to cross over each other. When the candidates reach each other, we simply stop simulating, and pretend that both candidates will from then stay at the position at which they met. We might attempt to allow $L$ to cross over $R$. Since $L$, as soon as they cross over $R$, becomes
    the more right-leaning candidate, we might define
    $$
    S_L(l,r) = \left\{ \begin{array}{cl} 
    \int_{-\infty}^{(l+r)/2} f(x) dx & \mbox{if $l <r$}, \\
    ~ & ~ \\
    \int_{(l+r)/2}^\infty f(x) dx & \mbox{if $l >r$}, 
    \end{array}
    \right.
    $$
    and the definition for $l=r$ would be a bit unclear. Similarly
    we would define $S_R$ distinguishing the two cases $l<r$ and $l>r$.
    Can you make sense of this mathematically? Does it make sense politically?
\end{question}

\subsection{Suggested homework questions}

\begin{probl} 
Suppose $f$ is as in Figure \ref{fig:ex_voter_beliefs}.
    Let $l=-1$ and $r=1.5$ initially, 
    and assume that $\alpha = \beta=0$, so 
    the candidates don't change positions with time. Who wins? 
\end{probl}

\begin{probl} 
In the previous problem, let $\alpha=0$, $\beta=1$. When the 
candidates meet each other and therefore stop changing positions,
who wins?
\end{probl}

\begin{probl} 
In the previous problem, let $\alpha=1$, $\beta=1$. When the 
candidates meet each other and therefore stop changing positions,
who wins?
\label{prob:alpha_beta_1}
\end{probl}

\begin{probl} 
In Figure \ref{fig:cand_moving_steep_ascent}, do $l$
and $r$ meet each other {\em tangentially} or at a non-zero angle? 
That is, is $\frac{d l}{dt} = \frac{dr}{dt} = 0$ at the time when
$l$ and $r$ meet, or not? (Answer this question using mathematics,
not just looking at the plot.) 
\label{prob:not_tangential}
\end{probl}

\begin{probl} 
\label{prob:median}
Prove: If $\alpha=\beta=0$, so the candidate positions don't move,
the winner is the candidate whose position is closer to the median. 
(The median of the voter belief distribution is the number $m$ with 
$
\int_{-\infty}^m f(x) dx = \frac{1}{2}.
$) This is a very simple special case of the {\em median voter theorem}.
\end{probl}

\begin{probl} (Programming problem) Try out what happens with 
the definitions (A.1) and (A.2) in Appendix A when the two 
candidates collide.
\end{probl}

\begin{probl} 
\label{prob:would_change}
Would the answer to problem \ref{prob:not_tangential} change if
(A.1) and (A.2) in Appendix A were the definitions of $S_L$ and
$S_R$?
\end{probl}

\section{When Voting is Not Mandatory}\label{sec:optional_voting}

In the United States, voting is not mandatory. If a voter doesn't feel strongly about either candidate they might choose to stay home. Therefore, it's not always in a candidate's best interest to be overly opportunistic, since they  might risk alienating their base. To account for this, we'll add a measure of {\it voter loyalty} to our model. In the voting literature, voter alienation is a more common term; see \cite{Leppel_09} but also \cite{Black_1948} and \cite{Hotelling_1929}. A high level of voter loyalty means that a voter is likely to stick with a candidate even as their position drifts, as long as that candidate is still the candidate closest to the voter's position.

We denote by $g(z)$ the probability
that a voter will still go vote if the candidate closest to them 
differs from their views by $z$. We assume
\begin{align}
    g(z)=e^{-z/\gamma}
\end{align}
where $\gamma>0$ is the model parameter measuring voter loyalty. 
Other choices of $g$ would be possible, but we always take $g$ to be
decreasing, differentiable, with $g(0)=1$.
We express the left and right candidate share of votes as a function of position as

\begin{align}\label{eq:SL_abstention}
    S_L=S_L(l,r)=\int_{-\infty}^{\frac{l+r}{2}}f(x) g(|l-x|)dx
    ~~\mbox{and}~~
     S_R=S_R(l,r)=\int_{\frac{l+r}{2}}^{\infty}f(x) g(|r-x|)dx.
\end{align}

\noindent
Section \ref{sec:mandatory_voting} is about the special case $\gamma=\infty$.

\subsection{Static Candidates}

\begin{figure}
\centering
  \includegraphics[width=5in]{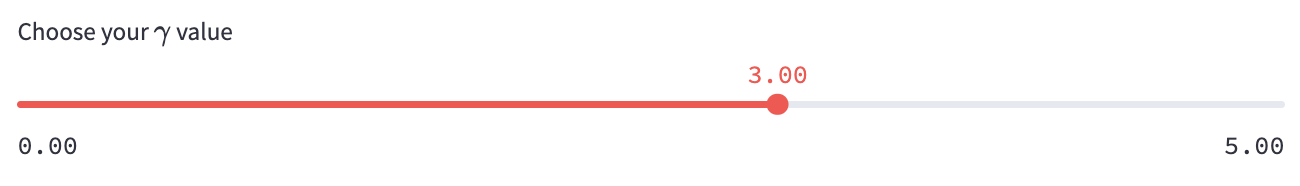}
  \caption{Slider with the set value of $\gamma=3$}
  \label{fig:example_gamma}
\end{figure}

\begin{figure}
\centering
  \includegraphics[width=5in]{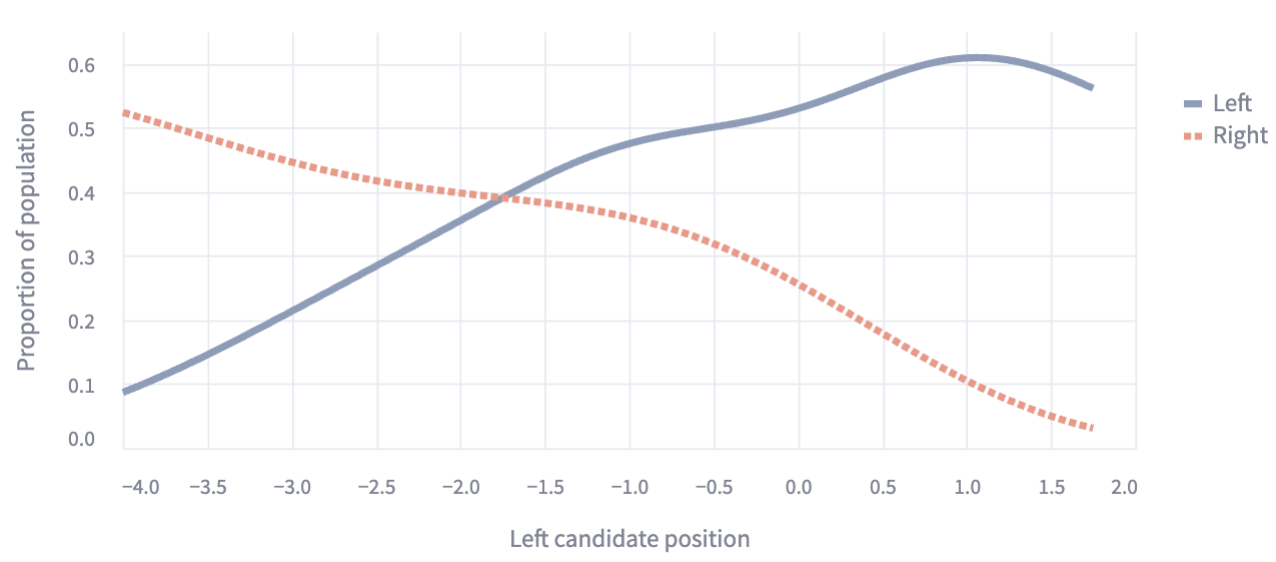}
  \caption{Proportion of population voting for each candidate as a function of position with voter loyalty}
  \label{fig:prop_voting_function_loyalty}
\end{figure}

Using the slider in our app, you can pick a value for $\gamma$. Remember, a greater value of $\gamma$ means that voters
are more {\em loyal}, i.e., more likely to stick with their candidate even as the candidate moves away from them.
We will go with the default value $\gamma=3$; see Figure \ref{fig:example_gamma}. In Figure \ref{fig:prop_voting_function_loyalty} we see the proportion of the population voting for each candidate as a function of position with finite voter loyalty.

\begin{question}
\label{summ_no_longer}
    Do the proportions in Figure \ref{fig:prop_voting_function_loyalty} always sum to 1? 
    \end{question}

\begin{question}
\label{moderates_win} 
    If the right candidate starts at 1.75, where does the left candidate have to be on the spectrum in order to win the election? Why?
\end{question}

\begin{question}
\label{loser_wins}
    Why is it possible for the left candidate to win with less that $50\%$ of the vote?
\end{question}

\subsection{Dynamic Candidates}

\begin{figure}
\centering
  \includegraphics[width=5in]{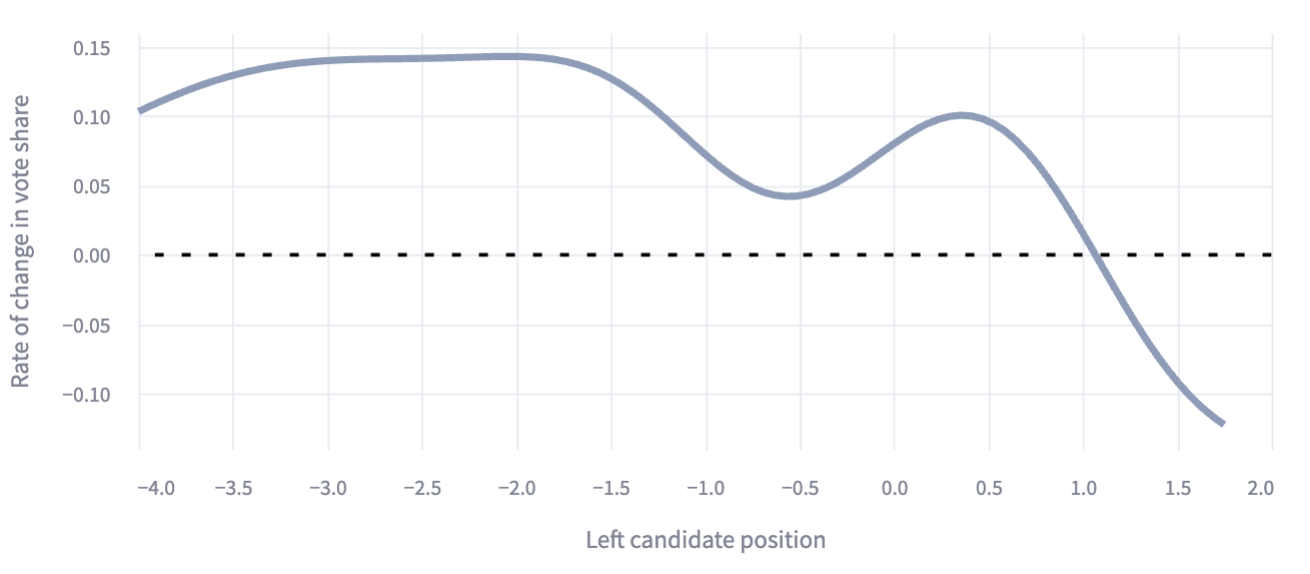}
  \caption{Rate of change in left candidate vote share, $\frac{\partial S_L}{\partial l}$, as a function of position when $\gamma=3.0$}
  \label{fig:rate_of_change}
\end{figure}

Next, we will look at how the share of votes changes as a function of candidate position, with the introduction of finite voter loyalty, assuming again equations \ref{eq:steepest_ascent}.
We see in Figure \ref{fig:rate_of_change} the rate of change in
the left candidate vote share as a function of position.

\begin{question}
\label{below_0} 
    What does it mean for points on this graph to be below zero?
\end{question}

\begin{question}
\label{where_are_they} 
    Looking at this graph, where are the fixed points and which of these are stable?
\end{question}

\subsection{Homework problems}

\begin{probl}
\label{prob:where_to_go}
Assume the distribution of Figure \ref{fig:ex_voter_beliefs}. 
Assume that the candidates are equally opportunistic, and that
people easily abandon their candidates, so $\gamma$ is small; try
$\gamma=0.2$ for instance. Suppose
that the right candidate is positioned at $r=1$. Where should the left candidate position themselves in order to win the election?

\end{probl}

\begin{probl} (Programming problem)\label{pr:tedious-problem2}
Use our Github code for this problem. Pick a different function $g$ that still makes intuitive sense. Then, implement all the steps we have covered so far in the app and compare with  $g(z)=e^{-z/\gamma}$. 
\end{probl}

\begin{probl}
\label{prob:dSL}
    Assume that $S_L$ is defined as in equation (\ref{eq:SL_abstention}). Show that for all $l$ and $r$, $l<r$
    \begin{align}
        \frac{\partial S_L}{\partial l}(l,r)=-\frac{1}{2}f\Big(\frac{l+r}{2}\Big)g\Big(\frac{r-l}{2}\Big)+\int_{-\infty}^{(l+r)/2}f'(x)g(|l-x|)dx
        \label{eq:dSL}
    \end{align}
The point is that the derivatives that appear in (\ref{eq:steepest_ascent}) can be evaluated analytically. (Of course,
the integral in (\ref{eq:dSL}) must still be evaluated numerically in the numerical solution of (\ref{eq:steepest_ascent}).) 

\vskip 5pt
\noindent
Hints: 
Break the integral in (\ref{eq:SL_abstention}) into two pieces, 
one from $-\infty$ to $l$ and one from $l$ to $(l+r)/2$.
Use
the fundamental theorem of calculus, the 
chain rule, and integration by parts. 
Assume that $g(0)=1$, $g$ is differentiable and decreasing,
and $f(x) \rightarrow 0$ as $x \rightarrow -\infty$. 
\end{probl}

\begin{probl} (Real life examples)
    Can you think of political candidates who have appeared to
    adjust their positions to enhance their chances of winning
    elections?
\end{probl}

\begin{probl} (Public Policy)
\label{prob:public_policy}
    What are some public policy initiatives that could increase voter participation?
\end{probl}

\section{Discontinuities}\label{sec:discountinuities}

\begin{figure}
\centering
  \includegraphics[width=5in]{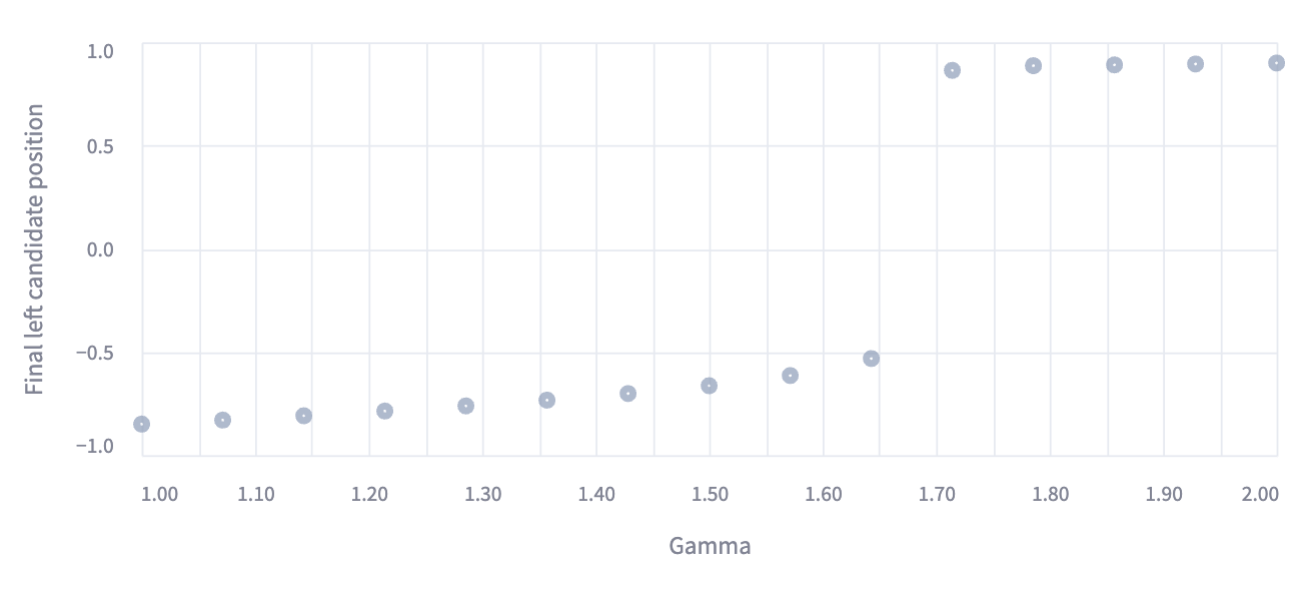}
  \caption{Left candidate's final position as a function of gamma}
  \label{fig:left_function_gamma}
\end{figure}

Changes in the political environment can have a 
discontinuous impact on candidate strategy. In Figure \ref{fig:left_function_gamma}, we illustrate this 
point with an example: Here 
$L$ starts at $l=-1$ and moves, but $R$ sits still at $R=2$.
The voter distribution is that of Figure \ref{fig:ex_voter_beliefs}.
We plot the final position of $L$ as a function of $\gamma$.
As you can see in Figure \ref{fig:left_function_gamma}, slight changes in $\gamma$ can cause abrupt substantial changes in candidate movements \cite{Borgers_et_al_candidate_dynamics}.

\begin{figure}
\centering
  \includegraphics[width=5in]{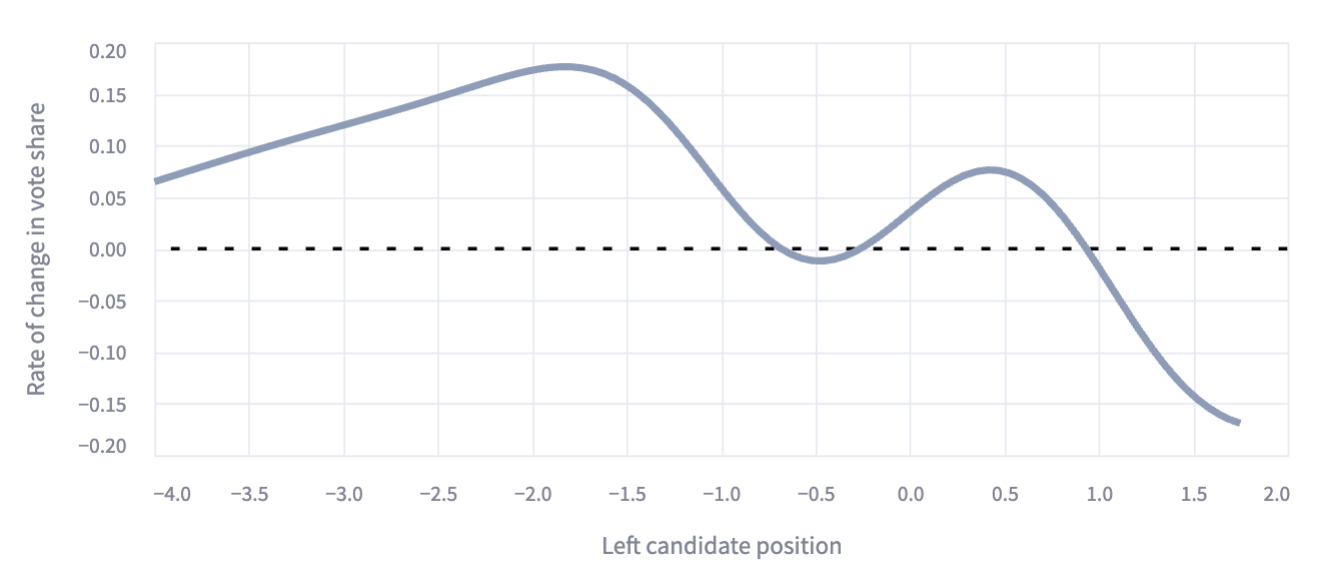}
  \caption{Rate of change in left candidate vote share as a function of position with $\gamma=1.7$}
  \label{fig:discontinuity_example}
\end{figure}

To understand how this discontinuity arises, look at Figures \ref{fig:rate_of_change} (where $\gamma=3$) and \ref{fig:discontinuity_example} (where $\gamma=1.7$). As $\gamma$ rises, the ``knee" in Figure \ref{fig:discontinuity_example} rises, and when it rises above the horizontal axis, the final position of $L$ 
abruptly jumps to the fixed point on the right. 
This is an example of a {\em saddle-node bifurcation} \cite{Abraham_1988,Strogatz_2015}. In a saddle-node bifurcation,
two fixed points collide as a parameter is changed, and annihilate 
each other. Thinking of the parameter change in the opposite direction,
one might also say that two fixed points appear ``out of the blue sky", 
and the bifurcation is then also called a {\em blue sky bifurcation}. 

If you want to try to generate this graph with different values, check out the  script {\tt optimal\_left\_position.py} in our Github repository.

\subsection{Homework problem}

\begin{probl}\label{prob:blue_sky}
    To illustrate how saddle-node bifurcations  give rise to discontinuities, consider the simple example 
\begin{equation}
\label{eq:blue_sky}
\frac{dx}{dt} = (x^2-\gamma)(1-x)
\end{equation}
where $\gamma \in (-1,1)$ is a parameter.
(a) Plot the right-hand side of (\ref{eq:blue_sky}) as a function
of $x$ for $\gamma=-0.5$, $\gamma=-0.1$, $\gamma=0$, $\gamma=0.1$, $\gamma=0.5$. 
(b) Let $x(0) = -1$. Compute and plot $x_\infty = \lim_{x \rightarrow \infty} x(t)$
as a function of $\gamma$. (Hint: No calculations to speak of are needed here.)
\end{probl}

\begin{probl} (Programming problem)\label{prob:unimodal_disc}
Start by reproducing our Figure \ref{fig:left_function_gamma}. Then, run the code with a unimodal $f$. How does the left candidate's final position as a function of gamma change now? Can you give an intuitive
explanation of the result?
\end{probl}

\section{Discussion}\label{sec:conclusion}

Mandatory voting would raise $\gamma$. Our model suggests that in a two-party system,
that may incentivize politicians to adopt centrist, compromising positions, in agreement
with \cite{Singh_2019}. However, it would surely affect $f$ as well, since with voluntary voting, some
population groups tend to participate less than others \cite{Fowler_13}. We note that $\gamma$ and $f$ can also be
affected by opening hours of polling stations, get-out-the-vote efforts \cite{Enos2014}, making Election Day a public holiday, etc. Mathematics cannot tell us whether such public policy measures are
desirable, but it  shed light on their potential consequences.

\section{Acknowledgment}

We thank the Tufts Data Intensive Studies Center (DISC) for financial support, and
the Tufts Institute for Research on Learning and Instruction (IRLI)  for helpful feedback on this project.

\vfill
\pagebreak

\begin{center}
{\Large Appendix A: Some thoughts on the discussion questions} 
\end{center}

\vskip 10pt
\noindent
{\bf Discussion Question \ref{reasonable}.} 
    Although US politics are often said to be deeply polarized, the distribution of ideological views on the left-right spectrum may still best be viewed as unimodal \citep{fowler2023moderates}.  Most people are ideologically moderate and hold a mix of liberal and conservative views. Ideology within each major party is approximately normally distributed, where the Democratic party has a more liberal mean than the Republican party, but there is a fair amount of overlap between the two distributions.  
        
    Nonetheless, many analysts are interested in how polarization affects political systems, and how candidates would respond to an electorate, or a group of potential donors, that was more polarized than the US public.

\vskip 10pt
\noindent
{\bf Discussion Question \ref{assump}.} 
    The model assumes that everyone in the population of eligible voters actually casts a ballot if they are required to by law, and that they vote for the candidate who has an ideological position most similar to their own position. 
  
  In fact, a much higher proportion of the population tends to vote in countries where voting is mandatory, but not $100\%$. Some mandatory voting laws are not enforced, and many penalties for not voting are minimal. People can also typically avoid penalties by casting blank ballots if they prefer not to choose a candidate. Voters do typically select candidates that are closer to them ideologically, but may have reasons for preferring a different candidate. In general, it is useful to think about what would happen in a simple setting even if it does not perfectly describe the real world.

\vskip 10pt
\noindent
{\bf Discussion Question \ref{why_so_close}.} 
    The left candidate can maximize their votes by moving just to the left of the right candidate, around 1.74 if the right candidate starts at 1.75.  They would win the support of all voters to the left of 1.74, but still not that of the voters to the right of 1.75. 

\vskip 10pt
\noindent
{\bf Discussion Question \ref{summ}.} 
    They do. When voting is mandatory, we assume that everyone votes for one of the two candidates.
    
\vskip 10pt
\noindent
{\bf Discussion Question \ref{where_to_go}.} 
    If the distribution of voter views is symmetric and the right candidate is fixed at position 1.75, the left candidate will winas soon as they position themselves to the 
    right of -1.75. They just need to be slightly more moderate than the right candidate.

\vskip 10pt
\noindent
{\bf Discussion Question \ref{non_symmetry}.} 
    If the left candidate is opportunistic, they are willing to move faster toward positions that will get them more votes. They will quickly start adopting more conservative positions. This will allow them to win the election, but they will end up adopting conservative positions to do so. (They will adopt more conservative positions than they would need to adopt in order to win.)

\vskip 10pt
\noindent
{\bf Discussion Question \ref{less_eager_wins}.} 
    The less eager candidate can only win if they start off at a more moderate position closer to the center. Even then they might lose if their opponent is sufficiently opportunistic and not too much more ideologically extreme.

\vskip 10pt
\noindent
{\bf Discussion Question \ref{what_if_they_cross}.} 
    $S_L$ and $S_R$ would now in general be discontinuous at $l=r$, which makes the meaning of the partial derivatives in (\ref{eq:steepest_ascent}) questionable. 
    If a politician very slightly to the left of their opponent shifts a bit to the right and is
    now very slightly to the right of their opponent, the voters 
    almost certainly would {\em not}  suddenly all swap their allegiances. Faced with two candidates of very similar views, the voters will not decide based on who is closest to their views any more, but based on other factors --- who gives the more rousing speeches, who has the better haircut, and so on.

    One might overcome some of these issues
    by defining $S_L$ and $S_R$ like this:
    \begin{eqnarray*}
    \nonumber
    \hskip 89pt
    S_L(l,r) &=& \int_{-\infty}^\infty f(x) s((x-l)^2-(x-r)^2) dx, \hskip 89pt(A.1)  \\
     \nonumber
    \hskip 89pt
    S_R(l,r) &=& \int_{-\infty}^\infty f(x) s((x-r)^2-(x-l)^2) dx,
    \hskip 89pt(A.2)
 \end{eqnarray*}
    where $s(z)$, $z \in \mathbb{R}$, is a function that is about 1 for $z<0$ and about $0$ for $z>0$, but transitions from 1 to 0 smoothly, say with $s(0)=1/2$. Now (\ref{eq:steepest_ascent}) 
    is defined for all $l$ and $r$. 
    
\vskip 10pt
\noindent
{\bf Discussion Question \ref{summ_no_longer}.} 
    They do not, since not everybody votes.
    
\vskip 10pt
\noindent
{\bf Discussion Question \ref{moderates_win}.} 
    With a symmetric distribution of ideology in the population and an equal level of loyalty for liberal and conservative members of the population, the left candidate still needs to cross -1.75 in order to win.

\vskip 10pt
\noindent
{\bf Discussion Question \ref{loser_wins}.} 
    Candidates only need a majority of votes cast,  not the votes
    of a majority of eligible voters. If many voters abstain, candidate don’t need a very large share of eligible voters
    to win.

\vskip 10pt
\noindent
{\bf Discussion Question \ref{below_0}.} 
    The candidate is losing votes as they move to the right.

\vskip 10pt
\noindent
{\bf Discussion Question \ref{where_are_they}.} 
    Fixed points are points at which the graph crosses the dashed line. Fixed points are stable if the graph has a negative slope (moving from positive values to negative values when going left to right) at the fixed point. Moving a bit to the left, the candidate would prefer to come back to the fixed point because they have more votes to gain. Moving a bit to the right, the candidate would start to lose votes and would want to come back to the fixed point.

\vskip 40pt

\begin{center}
{\Large Appendix B: Sketches of solutions for selected problems} 
\end{center}

Of course one learns  by thinking things through, not by
reading other people's answers. We don't want to spoil that, but 
we do want to 
help a reader who feels that they need a bit of help to get started. We therefore sketch
answers to some of the homework questions here; don't read our sketches unless you
feel you need to.

\vskip 10pt
\noindent
{\bf Problem \ref{prob:alpha_beta_1}.} $L$ will win. The reason is that
the situation is completely symmetric, except that $L$ has the
advantage of starting closer to the median of the distribution (which
is $0$ in Figure \ref{fig:ex_voter_beliefs}) than $R$. Can 
you state this argument more precisely?

\vskip 10pt
\noindent
{\bf Problem \ref{prob:not_tangential}.} They meet non-tangentially.
Think about what are $\frac{dl}{dt}$ and $\frac{dr}{dt}$ at the moment when $l$ and $r$
meet.

\vskip 10pt
\noindent
{\bf Problem \ref{prob:median}.} Let $m$ be the median of the distribution. If $m<(l+r)/2$, then 
$$
S_L = \int_{-\infty}^{(l+r)/2} f(x) dx > \int_{-\infty}^m f(x) dx 
= \frac{1}{2}.
$$
(We assume $f(x)>0$ everywhere, which is the case for a Gaussian mixture.) So if $m < (l+r)/2$, then $L$ wins. Similarly if $m>(l+r)/2$, then $r$ wins. Now convince yourself that $m<(l+r)/2$ means precisely the 
same as $|m-l| < |m-r|$, and $m>(l+r)/2$ means $|m-r| < |m-l|$.

\vskip 10pt
\noindent
{\bf Problem \ref{prob:would_change}.} The derivatives 
$\frac{\partial S_L}{\partial l}$ and $\frac{\partial S_R}{\partial r}$ are
$$
\frac{\partial S_L}{\partial l} = -2\int_{-\infty}^\infty f(x) s'((x-l)^2-(x-r)^2) (x-l) dx $$
and 
$$
\frac{\partial S_R}{\partial r} = -2 \int_{-\infty}^\infty f(x) s'((x-l)^2-(x-r)^2) (x-r) dx. $$
This implies (make sure you see why) that for $l=r$, 
$$
\frac{\partial S_L}{\partial l} = \frac{\partial S_R}{\partial r} =-2s'(0)(\mu-l) = -2s'(0)(\mu-r),
$$
where $\mu = \int_{-\infty}^\infty f(x) x dx$ is the mean voter
position. What does this imply about the question? Do equations
(\ref{eq:steepest_ascent}) drive $l$ and $r$ away from $\mu$, or towards
$\mu$?

\vskip 10pt
\noindent
{\bf Problem \ref{prob:where_to_go}.} Think about which votes $L$
will get when $l=-1$, and which votes $L$ will get when $l$ is nearly $1$. Why is it better for $l$ to be positioned at $-1$ than at $1$?

\vskip 10pt
\noindent
{\bf Problem \ref{prob:dSL}.} This is a somewhat
sophisticated calculus exercise. Here are some general calculus
formulas that are useful for it:
$$
\frac{d}{dl} \int_a^b h(x,l) dx = 
\int_a^b \frac{\partial h}{\partial l}(x,l) dx
$$
(differentiation under the integral sign), 
$$
\frac{d}{dl} \int_{a(l)}^{b(l)} h(x) dx = h(b(l)) b'(l) - h(a(l)) a'(l)
$$
(this follows from the fundamental theorem of calculus and the
chain rule), and the combination of these two:
$$
\frac{d}{dl} \int_{a(l)}^{b(l)} h(x,l) dx = h(b(l)) b'(l) -  h(a(l)) a'(l) + \int_{a(l)}^{b(l)} \frac{\partial h}{\partial l}(x,l) dx.
$$

\vskip 10pt
\noindent
{\bf Problem \ref{prob:public_policy}.} How about making Election Day a Sunday?

\vskip 10pt
\noindent
{\bf Problem \ref{prob:blue_sky}.} (b) $x_\infty=1$ if $-1<\gamma<0$, and $x_\infty=-\sqrt{\gamma}$ if $0 \leq \gamma < 1$. So the dependence of $x_\infty$ on $\gamma$ is discontinuous at $\gamma=0$.
\vskip 10pt
\noindent
{\bf Problem \ref{prob:unimodal_disc}.} There is no discontinuity with the unimodal $f$. The intuitive
explanation is as follows. Suppose first the distribution is bimodal.
There is therefore a substantial ``left-wing camp". As $L$ moves to the right, they lose left-wing votes, but gain right-wing votes. 
For large $\gamma$, the gain always outweighs the loss until $l$
reaches $r$. As
$\gamma$ drops, a window of values of $l$ is suddenly created in which 
$L$ is best off moving to the left to recover some of the lost left-wing votes. This is the reason for the discontinuity. It requires 
a substantial ``left-wing camp" of voters, in other words, a bimodal
distribution.

\end{document}